\documentclass[12pt]{iopart}
\usepackage{epsfig}

\newcommand{\Fig}[1]{figure~\ref{#1}}

\begin{document}

\title{New efficient methods to calculate watersheds}

\author{E~Fehr$^{1}$, J S~Andrade Jr.$^{1,2}$, S D~da~Cunha$^{3}$, 
L R~da~Silva$^{3}$, H J~Herrmann$^{1,2}$, D~Kadau$^{1}$, 
C F~Moukarzel$^{1,4}$, and E A~Oliveira$^{2}$}

\address{$^{1}$IfB, HIF E12, ETH Z\"urich, 8093 Z\"urich, Switzerland}
\address{$^{2}$Departamento de F\'{\i}sica, Universidade Federal do Cear\'a, 60451-970 Fortaleza, Cear\'a, Brazil}
\address{$^{3}$Departamento de F\'{\i}sica, Te\'orica e Experimental, Universidade Federal do Rio Grande do Norte, 59072-970 Natal-RN, Brazil}
\address{$^{4}$CINVESTAV del IPN Unidad M\'erida, Departamento de F\'{\i}sica Aplicada, 97310 M\'erida, Yucat\'an, Mexico}
\ead{ericfehr@ethz.ch}

\begin{abstract}
We present an advanced algorithm for the determination of watershed
lines on Digital Elevation Models (DEMs), which is based on the
iterative application of Invasion Percolation (IIP) . The main
advantage of our method over previosly proposed ones is that it has a
sub-linear time-complexity.  This enables us to process systems
comprised of up to $10^8$ sites in a few cpu seconds. Using our
algorithm we are able to demonstrate, 
convincingly and with high accuracy, the fractal character of watershed lines.

We find the fractal dimension of watersheds to be $D_f = 1.211 \pm
0.001$ for artificial landscapes, $D_f = 1.10 \pm 0.01$ for the Alpes
and $D_f = 1.11 \pm 0.01$ for the Himalaya.

\end{abstract}
\maketitle 

\section{Introduction}
The concept of watershed arises naturally in the field of
Geomorphology, where it plays a fundamental role in e.g. water
management~\cite{Vorosmarty98,Kwarteng00,Sarangi05},
landslide~\cite{Dhakal04,Pradhan06,Lee06,Lazzari06} and
flood prevention~\cite{Lee06,Burlando94,Yang07}. Moreover, important
applications can also be found in seemingly unrelated areas such as
Image Processing~\cite{Vincent91,Gao01,Shapiro01} and
Medicine~\cite{Bruyant02,Mao06,Vidal06,Yan06,Ikedo07}.
Watersheds divide adjacent water systems going into different seas and
have been used since ancient times to delimit boundaries. Border
disputes between countries, like for example the case of Argentina and
Chile~\cite{UN1902}, have shown that it is important to fully
understand the subtle geometrical properties of watersheds. 
Geographers and geomorphologists have studied
watersheds extensively in the past. There have also been preliminary
claims about fractality~\cite{Breyer92} but they were restricted to
small-scale observations and therefore inconclusive. Despite the far
reaching consequences of scaling properties on the hydrological and
political issues connected to watersheds no detailed numerical or
theoretical study has yet been performed. 

In particular in image processing or computer vision the interest in development
of efficient algorithms is high. There one tries to simplify and/or change an images 
representation such that it is more meaningful or easier to analyze.
This is done by image segmentation, e.g. partitioning a digital image into multiple
segments (sets of pixels, also known as superpixels). Typically one locates like
this objects and boundaries (lines, curves, etc.) in images. More precisely, image
segmentation is the process of assigning a label to every pixel in an image such
that pixels with the same label share certain visual characteristics as color, intensity
or texture. Adjacent segments are significantly different with respect to the same
characteristic(s) \cite{Shapiro01}. This sounds familiar and, although many other
methods, such as clustering, histograms \cite{Ohlander78}, edge-detection \cite{Pathegama05},
region growing \cite{Vincent91}, level-set or graph partitioning \cite{Shi00}, have been developed for that
purpose, also watersheds are currently used. This because, if one transforms an image
to its color gradient representation and calculates the watersheds on the so achieved DEM
like data, one can identify the obtained catchment basins as the different (color) regions and
their boundaries to be represented by the watersheds.

It is the purpose of the present paper to present an advanced method, which allows
us to study the geometrical properties of watersheds on huge landscapes and with
high statistics. Applied to an artificial landscape model we show results on length
scales spanning over three orders of magnitude yielding highly accurate estimates 
of the fractal dimension of watersheds. Our algorithm is based on an iterative
invasion-percolation (IP) like model. We compare to other algorithms in use and show
the advances of our method. Furthermore, we prove the equivalence of our IP-based method
to the so called flooding algorithm. Finally the method is applied to Digital Elevation Maps
(DEM) from regions including the Alps and the Himalayas.

\section{The two methods}
Traditional cartographical methods for basin delineation relied on
manual estimation from iso-elevation lines and required a good deal of
guess work. Modern procedures are based upon the automatic processing
of DEM or Grayscale Digital Images where gray intensity is transformed
into height. One of the most popular algorithms for watershed
determination~\cite{Vincent91} (compare also \emph{flooding} later) uses
rather complicated data structures and at least one pass over all pixels in
order to calculate watersheds, and is adequate for grayscale images,
i.e. integer-height spaces. In the following, we will present a much more
efficient technique.

Let us consider a DEM with sites $i$ having heights $h_{i}$. We define the set of
sites $S_k$, $k=0,1,\ldots, N_{sinks}$, to be the \emph{sinks} of our system. On natural
landscapes these sinks are the natural water-outlets of the terrain, as for
example a lake, a river or an ocean.
We consider that at each time step, water flows to the lowest-lying site on the
perimeter of a flooded region. If when starting from $i$, sink $S_k$ is flooded
before any other, then site $i$ and the whole flooded region are considered to belong
to the \emph{catchment basin} of sink $S_k$, equivalently one says that $i$
drains to $S_k$. This procedure introduces a classification of lattice sites, i.e.
a subdivision into non-overlapping subsets whose union is the entire lattice.
The sets of bonds separating neighboring basins we call \emph{watersheds}.
We are interested in the geometrical properties of a single watershed line.
But as natural landscapes normally contain more than one sink and hence more
than one single line, it is difficult to study long range properties. We therefore restrict
ourselves to study two-sink systems, or in other words studying only the main
watershed with respect to these two sinks. Although we use these restrictions, our algorithm,
which we will describe in the following, is able to deal also with multiple-sink
systems. This because it is not restricted to certain number of sinks/labels but
is free to deal with as many as one would like to.

\begin{figure}[!ht] 
\centerline{\epsfig{figure=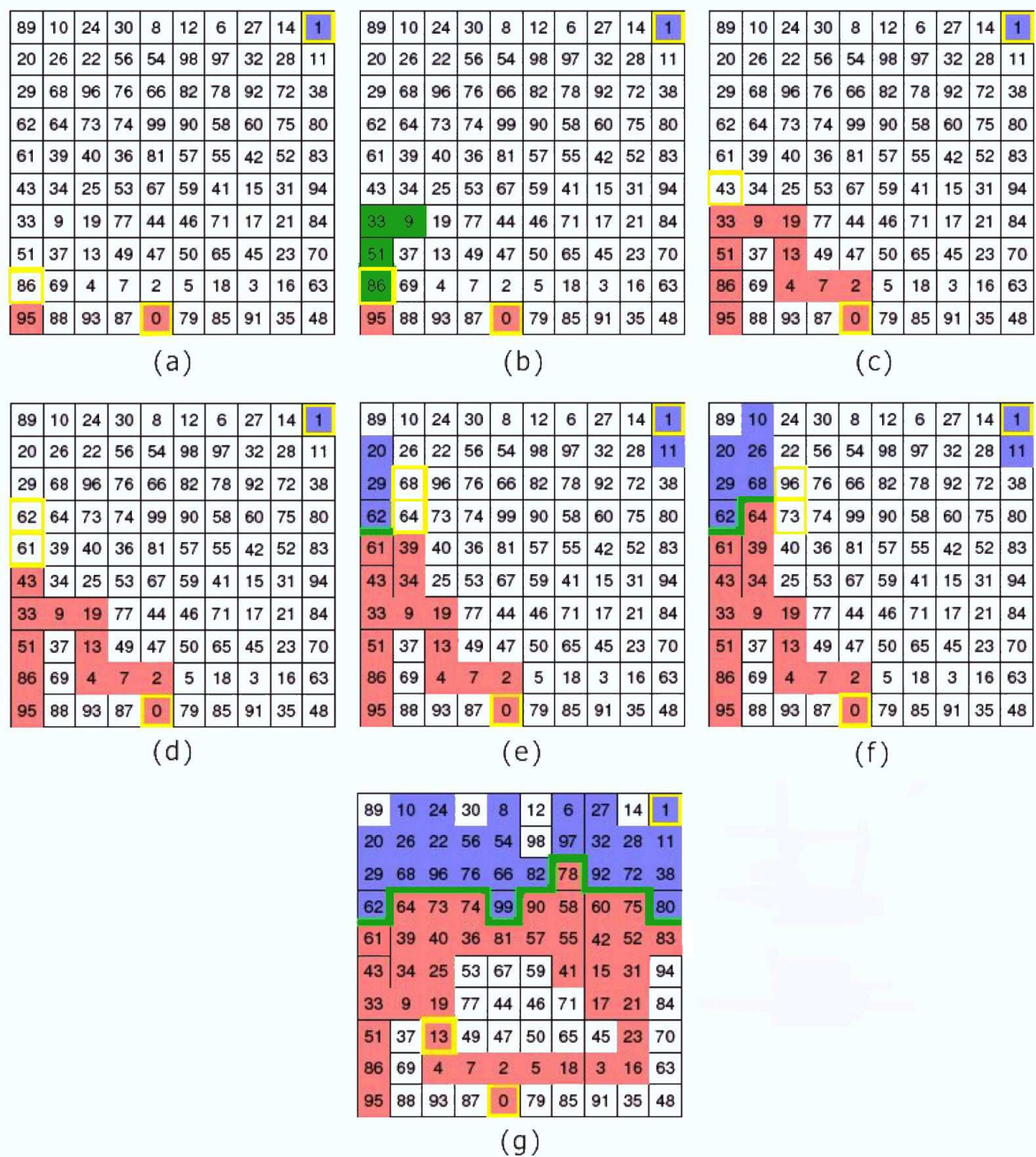,width=\textwidth}}
\caption{Different stages during the search for the main watershed (green lines), red/blue for the labels to the two sinks, green are visited but not yet labeled sites, the numbers mark the integer heights of the corresponding site and the yellow box surrounds the starting site of the current IP-cluster, right and left periodic boundary conditions are applied. Upper most and lower most bonds connect to the sinks (which are entire rows). Further the lowest two sites are already labeled. (a)-(c) shows the growth of an IP-cluster until reaching a sink. (a)-(e) search for the first watershed bond, growing IP-clusters of each site along the first column consecutively. (e)-(g) Following the watershed.}
\label{algo_how}
\end{figure}

\begin{figure}[!ht] 
\centerline{\epsfig{figure=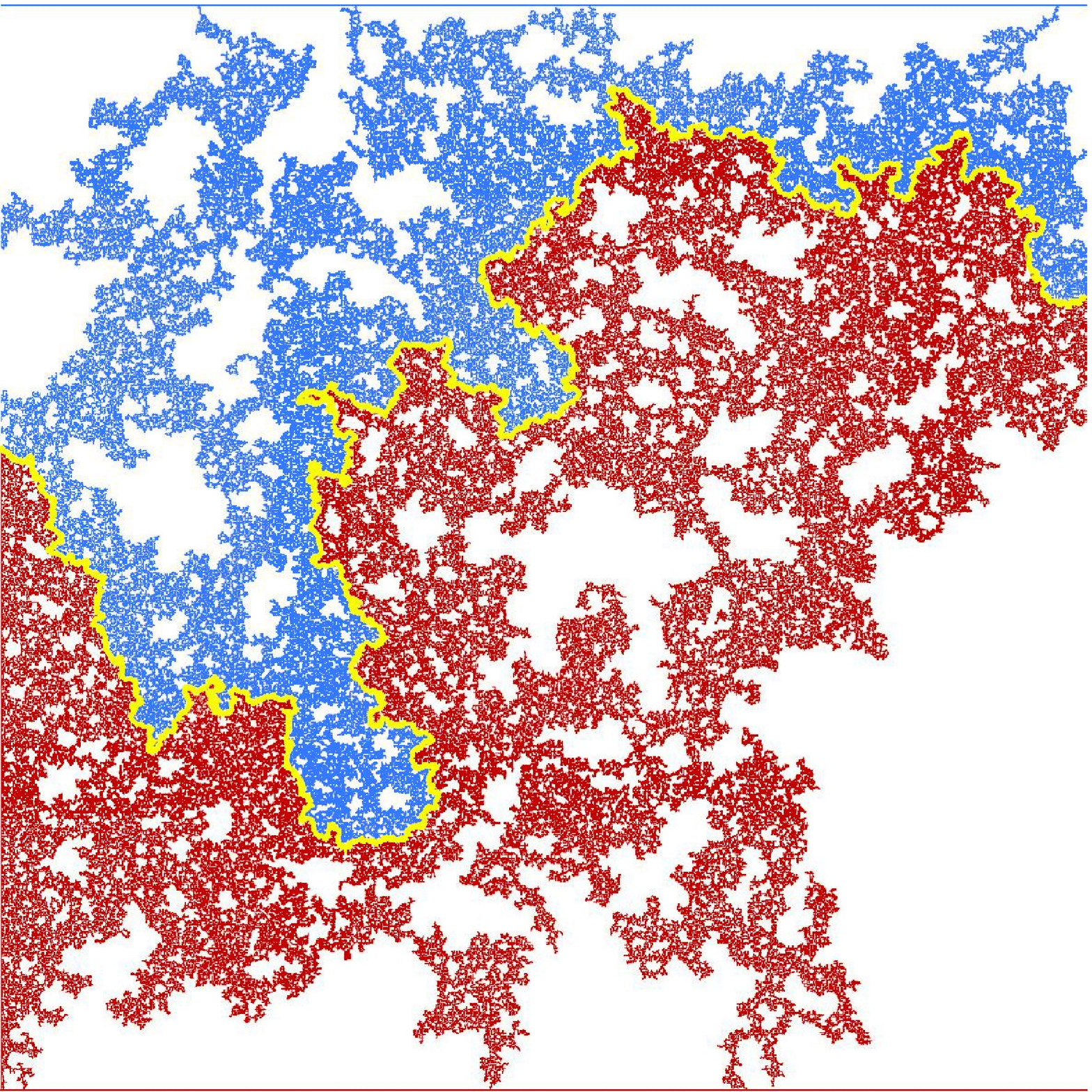,width=9cm}}
\caption{Labels of the sites for a random landscape obtained during the run of the IP-based algorithm. Red labels (lower part) belong to the sink at the bottom row and the blue ones (upper part) to the sink at the upper most row. The watershed is marked in yellow. Only the colored sites are visited by the algorithm, meaning that white sites have not to be considered by the algorithm in order to obtain the right set of watershed bonds.}
\label{algo_example}
\end{figure}

In our algorithm, that we call \emph{IP-based algorithm}, a cluster is
started from site $i$ and grown by adding, similar to Invasion
Percolation (IP), at each step, the smallest-height site on its
perimeter until the first sink is reached (see figure \ref{algo_how}). This can be done rather quickly when
using binary heaps or other tree methods to sort the list of perimeter sites according
to their height values. This procedure is related
to Prim's algorithm for growing Minimum Spanning Trees (MST) from
$i$~\cite{Dobrin01,Barabasi96}. When implemented blindly over the
whole lattice, however, this process would be inefficient since in
principle a new cluster has to be grown from each site $i$. By noting
that all sites occupied by a cluster at the time the sink is reached
also drain to that sink, it becomes clear that we can set for each site in the cluster
a link to that sink. This leads to two improvements, first, we do not need to treat again
the sites contained in the cluster as we can directly link them to the sink reached
by the cluster, and second, we can stop growing clusters already when we reach a site
which has an existing link to a sink. Like this, the algorithm has to visit each site
only once.
But still each site on the map has to be visited. Let us assume we already know a line dividing
the system and we want to check if this line is really the watershed of the current system. To prove this
we actually only need to test the sites adjacent to the watershed line if they belong to
different sinks on the two sides of the line. Hence the subset of the system, formed by
these adjacent sites, is the minimal subset we have to test, meaning to grow IP-clusters
from them, in order to clearly determine the watershed. Unfortunately we do not know in
advance the exact form or position of the watershed, neither we know any of the sites in
that minimal subset. The only thing we know is, that the watershed is located somewhere
in between the two sinks. This means that, whatever path we follow, to go from one sink to
the other, we have to cross the watershed. The sites we visited on such a path, before the watershed
traversal, belong to the sink we started from and the sites afterwards belong to the second
sink. This is what we now can use to find the first two sites of the minimal subset. We follow a
straight line of sites connecting the two sinks. Starting from one sink we follow the
connection site by site and grow IP-clusters from these sites. We proceed until we
reach the first site that drains to the second sink, which means
that we have crossed the watershed (see figure \ref{algo_how}). The bond between the last two sites, that drain
to different sinks, is part of the watershed line and the two sites belong to the minimal subset as
mentioned above. Now we can reconstruct the minimal subset and the set of bonds forming
the watershed line by just following step by step the direction of the already found watershed at
one of its two ends and testing the next two sites to find if the direction changes by $\pm 90$
degrees or remains the same. We just walk along the watershed and hence have only
to test slightly more sites than in the minimal subset (see figure \ref{algo_how}). Of course the IP-clusters for these sites
have to be grown until they reach a sink or sink-connected site, but there are typically large
parts of the terrain that need not be visited at all by the algorithm in order to determine the entire
divide. For use in multiple sink systems one would only have to consider additionally branching
and joining of the watershed lines (also only one starting bond has to be found).
That only a part of the system has to be considered (compare figure \ref{algo_example}) and there each site only once, is probably the biggest advantage of this procedure. The described algorithm is fast enough
to allow for the determination of watersheds on lattices comprising $10^8$ sites, in a
few CPU seconds on a normal workstation.

In the following we describe a slightly adjusted version of the commonly used
watershed algorithm for image segmentation (\cite{Vincent91}) in order to compare
efficiency and prove equivalency to our IP-based algorithm (actually it is making use of 
disjoint-set data structures). In this procedure, which we call \emph{flooding}, the whole
lattice is flooded (or occupied) in order of increasing height, i.e. at time $t$ all
sites with $h_i<h(t)$ are occupied and $h(t)$ grows site per site in time. This procedure
is also related to Kruskal's algorithm for growing MST's~\cite{Dobrin01,Barabasi96}. 
It is assumed that each site has a height, which is a real number, such that after sorting them, a
unique sequence of sites is defined. At the beginning, each sink is assigned a
different label, and we will furthermore specify that a cluster of occupied sites that
gets connected (through a path of occupied sites) with a sink is labelled accordingly,
i.e. labels propagate from the sinks. Clusters of occupied sites not yet in contact with
any sink remain unlabeled. Whenever two different labels would get in contact with
each other because of the addition of a new site, the (unique) bridging site is first
labelled with the label of its lowest-lying neighbor. Next all bonds connecting
sites with different labels are ``cut'' so that labels from different sinks never mix.
Bonds connecting different labels are called \emph{bridges}. When this flooding
procedure is completed, the different labels identify the corresponding catchment
basins and the set of bridges, i.e. the set of bonds separating the different basins,
identifies the watershed(s). In the described form the sinks are predefined by hand
or applying some criteria. Of course one can reformulate this procedure to identify all
intrinsic catchment basins automatically by just introducing new labels when a site gets
occupied that has only non-occupied neighbors. This will in fact return all
possible watersheds on the current landscape, even of the smallest basins,
which is probably not what one would like to have.
Unfortunately using the above described flooding algorithm one has in general to
consider each site or pixel to completely determine the watershed, such that, already
without considering the needed sorting, this algorithm performs less efficient than
the presented IP-based algorithm.

\section{Equivalence of Methods}
We now sketch a simple proof for the equivalence of these two procedures. 
During the flooding growth a set of unlabeled sites that become labelled by occupying
site $b$ is called a \emph{lake} and $b$ is called its \emph{outlet}. The height of
all sites in a lake is bounded by that of its outlet, i.e. $h(i) \leq h(b(i))$. Let us stop the
flood procedure when site $i$ gets labelled for the first time, say with label $S_k$.
Now start an IP cluster from $i$ and let it grow for as long as it takes until some sink
is first occupied. During the growth of this cluster, which proceeds by always adding
the lowest-lying site on the perimeter, several lakes and their outlets may be occupied.
We need to demonstrate that no bridge will be ever traversed during this IP growth,
and therefore that sink $S_k$ will be occupied first. During the growth of the IP cluster,
a lake will be entirely flooded i.e. the water will reach the level of its outlet. Since all
bridges by definition are higher than the outlet of a lake (because by definition bridges
were occupied later than the outlet), clearly the next occupation will proceed through
the outlet and not through the bridge. Therefore no bridge will be ever traversed thus
proving the equivalence of both procedures.

\begin{figure}[!ht] 
\centerline{\epsfig{figure=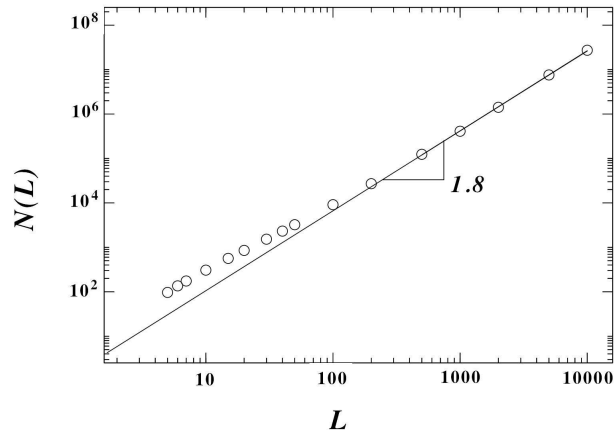,width=9cm}}
\caption{Log-log plot of the number $N(L)$ of sites visited by the IP-based algorithm, in order to determine the watersheds on artificial landscapes, as a function of linear dimension $L$ (circles). Solid line shows the least squares fit to the five last data points, which gives a dimension $D_f = 1.8 \pm 0.01$ for large scales ($N \sim L^{D_f}$).}
\label{IP-based scaling}
\end{figure}

It is important to note that, although conceptually equivalent, the IP and the flooding
algorithms are markedly different in terms of computational performance. First, the
flooding needs a sorting of the sites according to their heights, which needs at least
$O(N)$ time. Additionally it has been shown by Fredman and Saks in \cite{Fredman89}
that building the disjoint-set data structure, which the algorithm is based on, needs in all
cases at least $O(\alpha(N) N)$, where $\alpha(N)$ is the inverse Ackermann function,
which is constant for almost all values of $N$. Hence the flooding scales at least with $O(N)$.
In figure \ref{IP-based scaling}, for the random artificial landscape case, the number of sites visited by the IP-based algorithm is shown together with its scaling in linear dimension $L$, which clearly points out the sub-linear time-complexity of this method. The total number of visited sites $N$ scales as $N \sim L^{D_f}$ with $D_f = 1.8 \pm 0.01$. This is comparable to what is found in nature. There, river networks show fractal dimensions between $1.7$ and $1.9$ \cite{Rodriguez97,Maritan96,DeBartolo06,Horton32,Hack57,Melton58}. On large scales we expect this fractal dimension to be dominated by the largest IP-cluster we grow, for which we found a fractal dimension that is in good agreement with literature and the result for the total number of visited sites. Furthermore we expect that the algorithm is even more efficient on natural landscapes, as the largest IP-cluster will most probably follow the main stream. Hence scaling like the main stream, which is reported to have fractal dimensions between $1.0$ and $1.2$ \cite{Rodriguez97,Maritan96,DeBartolo06,Horton32,Hack57,Melton58}.

\section{Results}
\begin{figure}[!ht] 
\centerline{\epsfig{figure=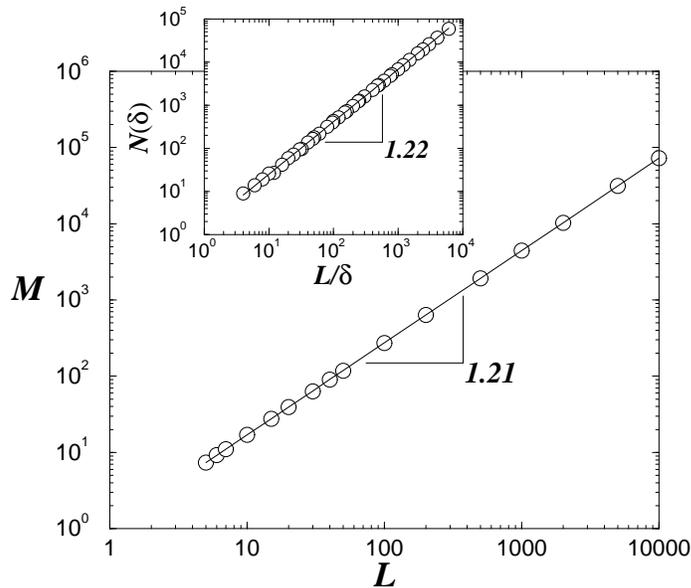,width=9cm}}
\caption{The mass $M$ of the watershed of (uncorrelated) random-height maps 
scales with linear dimension as $M \sim L^{D_{f}}$, with $D_{f}=1.211
\pm 0.001$. A box-counting procedure (inset) gives a consistent
result. This fractal dimension is independent of the type of disorder,
as long as each height is an uncorrelated random variable.}
\label{ws scaling}
\end{figure}

As an example we applied the IP-based algorithm to artificial landscapes., where the height
of each site on the lattice is an independent random variable uniformly distributed between
0 and 1~\footnote{Since the watershed location only depends on the order in which sites are occupied, it is clear that any distribution of heights will produce the same statistical results, as discussed for example in \cite{Dobrin01}.},
and two sinks are defined respectively as the upper- and lowermost
lines of the lattice. Due to the high efficiency of the method we could process a huge amount
of data and hence gather a lot of statistics. This enables us to very precisely estimate the fractal
dimension of the watershed using the mass-scaling. We measured the total length of the resulting
watersheds and averaged this value over at least $10^4$ samples for a given lattice size. The
results for artificial landscapes are shown in figure \ref{ws scaling}. Our data is strikingly straight on a log-log plot, which warrants neglecting finite size corrections. Hence, the fractal dimension is
measured as the least squares fit to the slope of the shown data. We find that the watershed
is a fractal, i.e. $M \sim L^{D_f}$, with a fractal dimension $D_f=1.211 \pm 0.001$~\footnote{We repeated the mass-scaling analysis using the box-counting method on individual samples of $L=1.5 \times 10^{4}$, and obtained a consistent result $D_f=1.21 \pm 0.01$.}.
This value of $D_f$ is close to that found for Disordered Polymers ($\approx 1.2$~\cite{Cieplak94}),
``strands'' in Invasion Percolation ($1.22 \pm 0.01$~\cite{Cieplak96}),
and paths on MST's ($1.22 \pm 0.01$~\cite{Dobrin01}).
The roughness exponent found for the watersheds is equal
to unity within the statistical error bars, supporting the fact that they are indeed self-similar fractal
objects and not self-affine.

\begin{figure}[!ht] 
\centerline{\epsfig{figure=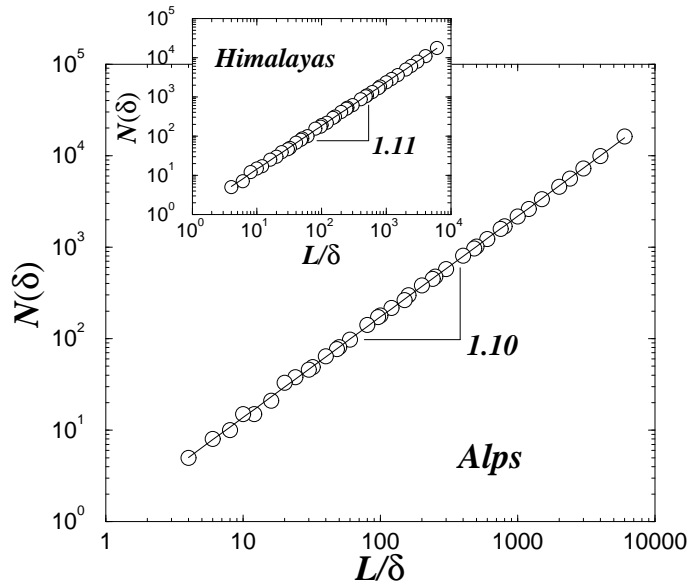,width=9cm}}
\caption{Log-log plot showing the box-counting results for the Alps watershed. 
The solid line is the best fit to the data which gives a fractal
dimension of $D_{f}^{Al} = 1.10 \pm 0.01$. The inset shows our results
for the watershed calculated on the Himalayas. The fractal dimension
in this case is $D_{f}^{Hi} = 1.11 \pm 0.01$.}
\label{ws natural}
\end{figure}

We also applied the IP algorithm to study the fractal properties of
watersheds on naturally occurring landscapes, which clearly have long
range correlations \cite{Turcotte97}. Of course, if we would like to estimate the fractal dimension
of a single watershed line of one given landscape, as is the case for these natural
topographies, we cannot work with the statistic approach used above for the artificial
ones. Therefore we have to choose another method. As we follow the watershed bond
per bond and hence have the correct ordered set of bonds, we can easily apply a
yard stick method. Measuring now the length of the watershed in terms of a given
yard stick size, meaning measuring the length on a certain resolution, we can
define the corresponding fractal dimension again. We checked this procedure
also on several realizations of artificial landscapes and found the newly estimated
fractal dimension to agree with the above presented $D_f=1.211 \pm 0.001$ within
the error bars. Therefore we can assume that this equality is also true for natural
landscapes. Now, using topographic data derived from the Shuttle
Radar Topography Mission (SRTM)~\cite{Farr07}, we analyzed
several mountainous regions and determined their watersheds within
the available resolution of 3 arc-sec (i.e. roughly $90~m$). As shown in
\Fig{ws natural}, a yard-stick analysis of the watershed performed in
the range $100~m<L<200~km$ for the Alps and Himalayas gives fractal
dimensions that are indistinguishable within their error bars, namely
$D_{f}^{Al}=1.10 \pm 0.01$ and $D_{f}^{Hi}=1.11 \pm 0.01$, respectively.
The error bars were obtained by determining the lines of maximal and of
minimal slope that would still be consistent with the data in \Fig{ws natural}.
Again the roughness exponent found for the watersheds is equal to unity within
the statistical error bars, hence we have still self-similar fractal objects and not self-affine
for intermediate to large scales. The origin of the upper and lower-size scales is clear.
On large scales ($>200~km$) the watershed follows the direction of the main crust foldings, which
depends on processes occurring on the scale of tectonic plates, which are non-fractal.
Hence the scaling should be essentially linear ($D_f=1$). Although beyond our resolution,
at small scales ($<100~m$), below the size of an individual mountain, the watershed would
connect peaks and troughs, which are typically self-affine.

It is important to notice that the fractal dimension of the watershed
line $D_f$ is by definition independent of the type of disorder on the
landscape, as long as each height is an uncorrelated random
variable because only the spatial order of the random variables 
(heights) in the system matters and not their relative
numerical differences. The small discrepancies (around $10\%$) between
fractal dimensions of watersheds in the natural topographical data
taken from SRTM and the artificial random landscapes can be
explained by the presence of correlations in the first case. In
particular, long-range correlations in space have been reported in a
large variety of physical, biological and geological systems
\cite{Family91,Bunde94,Sahimi98,Sapoval04,Schorghofer02}.

\section{Conclusions}
We presented an advanced numerical algorithm of sub-linear time-complexity and showed
its equivalence to a currently used watershed algorithm. Giving some qualitative
arguments and analysis we pointed out the improved efficiency of the presented
method. We further investigated the watershed topology of natural and
synthetic DEM's. Our results show that watersheds generated on very
large ($10^{8}$ sites) uncorrelated random landscapes are self-similar
with fractal dimension $D_f=1.211 \pm 0.001$. Natural watersheds
calculated from landscapes of mountainous regions like the Alps and
Himalayas also display self-similarity, but with slightly smaller
fractal dimensions, $D_{f}^{Al}=1.10 \pm 0.01$ and $D_{f}^{Hi}=1.11
\pm 0.01$, respectively. The difference between the fractal dimensions
arises probably due to the fact that long-range correlations exist in the natural
system. The fractality of watersheds has widespread consequences on its
susceptibility to perturbations of the topology and transport properties at
the boundary of catchment basins. These are questions that will be investigated
in the future.

\ack
We acknowledge useful discussions with P.~Duxbury, M.~Alava,
H.~Seybold and A.~A.~Moreira. We also thank the agencies CONACYT
(48783,74598), M\'exico, and CAPES, CNPq and FUNCAP, Brazil, for
financial support.

\bibliographystyle{prsty}

\end{document}